\documentclass[epj]{svjour}
\usepackage[T1]{fontenc}
\usepackage[latin1]{inputenc}
\usepackage{ae,aecompl}
\usepackage[english]{babel}
\usepackage{psfrag}
\usepackage{amssymb}
\usepackage{thumbpdf}

\usepackage{color}
\definecolor{darkred}{rgb}{0.5,0,0}
\definecolor{darkgreen}{rgb}{0,0.5,0}
\definecolor{darkblue}{rgb}{0,0,0.5}

\newcommand{\D}{\displaystyle}

\makeatletter

\newif \ifpdf
    \ifx \pdfoutput \undefined
        \pdffalse
    \else
	\pdfoutput=1
        \pdftrue
\fi

\ifpdf
    \usepackage[pdftex]{graphicx}
    \pdfcatalog {
        /PageMode (/UseNone)
    }
    \usepackage[pdftex,
		colorlinks,
		bookmarksnumbered,
		bookmarksopen,
		pdfstartview=FitH,
		linkcolor=darkblue,
		filecolor=darkgreen,
		urlcolor=darkred,
		citecolor=darkblue
		]{hyperref}
		
\else
    \usepackage{graphicx}
    \usepackage[ps2pdf]{hyperref}
\fi

\makeatother

\begin{document}

\title{Thermodynamics of ionization and dissociation in hydrogen plasmas including fluctuations and magnetic fields}

\titlerunning{Thermodynamics of hydrogen plasmas}

\author{Werner Ebeling\thanks{\email{ebeling@physik.hu-berlin.de}}\and 
Hendrik Hache\thanks{\email{hache@physik.hu-berlin.de}} \and 
Michael Spahn\thanks{\email{spahn@physik.hu-berlin.de}}}

\institute{Institut für Physik, Humboldt-Universität zu Berlin, Invalidenstr.
110, 10115 Berlin, Germany}

\date{\today}

\abstract{Applying Gibb's geometrical methods to the thermodynamics of
H-plasmas we explore the landscape of the free energy as a function of the
degrees of ionization and dissociation. Several approximations for the free energy
are discussed. We show that in the region of partial ionization/dissociation the quantum Debye-Hückel approximation (QDHA) yields a rather good but still simple representation which allows to include magnetic field and fluctuation effects. By using relations of Onsager-Landau-type the probability of fluctuations and ionization/dissociation
processes are described. We show that the degrees of ionization/dissociation are probabilistic quantities which are subject to a relatively large dispersion. Magnetic field effects are studied.
}

\maketitle

\section{Introduction}

In plasmas and in particular in hydrogen systems, ionization and dissociation
processes play an important role. We find the equilibrium values of
the degrees of ionization and dissociation by minimization
of the free energy at fixed total mass \cite{ebkrkr76,ebetal91,beuleal99}.
Further the ionization and dissociation rates are derived from kinetic
equations \cite{ebetal91,klim}. Here we will start with 
a study of the geometry of the free energy landscape \cite{ebfree90,ebhi00}.
The idea to apply geometrical methods in thermodynamical research goes back to
pioneering papers of Gibbs \cite{gibbs1892}. We apply Gibb's method here to
plasmas. 
We will study different approximations to the thermodynamic functions and the 
influence of several effects, our approach is limited to the 
region $T > 10000 K$.
In a preceding work we have developed a similar approach for
ionization processes described by the Saha equation \cite{ebhi00}. We include
here also effects of magnetic fields which are of increasing interest not only
for astrophysical applications but also for plasmas confined in magnetic
traps. A preliminary work on magnetic field effects was published earlier
\cite{ebstor00}. 
The magnetic field introduces an anisotropy into the system. This changes the
ideal gas contribution, the bound state energies and the contributions due to
interactions of the free particles. Here we will take into account the ideal gas 
corrections in an exact way, and the bound state and interaction corrections in
approximations which include the quadratic orders in the field $O(B^2)$. 
Practically this restricts our approach to fields $B < 10^5 T$. 

We start with exploring the geometry of the free energy
landscape as a function of the degrees of ionization and dissociation. On this
basis we are able to describe not only the chemical equilibria but also
fluctuations around the equilibrium values \cite{ebfree90}.

\section{The free energy density of hydrogen}
\label{sec:ContribFreeE}

\subsection{Defining the basic variables of the system}

We consider in the following hydrogen plasmas at fixed temperature
\( T \), and proton density \( n \). Our aim is to study ionization
processes\begin{equation}
H\Leftrightarrow p+e^{-}
\end{equation}
and dissociation processes\begin{equation}
H_{2}\Leftrightarrow H+H.
\end{equation}
For simplicity, the formation of \( H_{2}^{+} \) and \( H^{-} \)
species will be neglected.

In order to define the thermodynamics of the system we introduce the
free energy
\begin{equation}
F\left( T,V,\left\{ N\right\} \right) ,
\end{equation}
the density of the free energy
\begin{equation}
f\left( T,n\right) = \frac{F}{V} = f\left( T,n_{e},n_{i},n_{a},n_{m}\right) ,
\end{equation}
the degrees of ionization and dissociation \cite{beuleal99}.
We will use here the following definitions:
\begin{equation}
\begin{array}{r@{\D \;=\;}l@{\quad}r@{\;=\;}l}
\rule[-4ex]{0mm}{8ex} \D \alpha & \D \frac{n_{i}}{n_{i}+n_{a}+2n_{m}} & \D \beta & \D \frac{n_{a}}{n_{a}+2n_{m}} \\ 
\rule[-4ex]{0mm}{8ex} \D \beta_a  & \D \frac{n_a}{n_i + n_a + 2 n_m} & \D \beta_m & \D \frac{2n_m}{n_i + n_a + 2n_m}.
\end{array}
\end{equation}
We note that $\beta_a$ is the relative amount of protons bound in atoms
and $\beta_m$ the relative amount bound in molecules.
Due to the balance relation for the 
total proton density
\begin{equation}
n=n_{i}+n_{a}+2n_{m}.
\end{equation}
we find the relations
\begin{equation}
\beta_a = (1-\alpha)\beta, \, \beta_m = (1-\alpha)(1-\beta).
\end{equation}
The condition of neutrality requires that electron and ion densities
are always equal:
\begin{equation}
n_{e}=n_{i}.
\end{equation}
Therefore the free energy density depends only on 4 independent parameters
\begin{equation}
f\Rightarrow f\left( T,n,\alpha ,\beta \right) .
\end{equation}
If we include magnetic field effects we have 5 independent parameters
\begin{equation}
f \Rightarrow f\left( T, n, B, \alpha ,\beta \right) .
\end{equation}
The equilibrium composition is found by the minimization procedure
\begin{equation}
\frac{\delta f}{\delta \alpha }=0,\quad \frac{\delta f}{\delta \beta }=0.
\end{equation}
Instead of $\beta$ we may use $\beta_a$ or $\beta_m$ as variational parameters. The relation 
\begin{equation}
\label{eq:sumabc}
\alpha+\beta_a+\beta_m=1
\end{equation}
gives the possibility for visualization of the free energy surface on a simplex.

\subsection{The different contributions to the free energy}

We assume the following structure \cite{ebetal91,beuleal99}:
\begin{equation}
f=f_{e}^{id}+f_{i}^{id}+f_{a}^{id}+f_{m}^{id}+f_{ee}+f_{ii}+f_{ie}+f_{hs}.
\end{equation}
The first 3 terms describe the ideal contributions of the electrons,
ions, atoms and molecules. The next 3 terms represent the Coulomb interactions electron-electron, ion-ion, 
and ion-electron. For these terms different approximations exists as e.g.\ the quantum Debye-Hückel-ap­prox­i­mation 
(QDHA) and Padé approximation \cite{ebkrkr76}. Basically we will use here the QDHA:
\begin{equation}
f_{ee}+f_{ii}+f_{ie}\approx f_{DH}.
\end{equation}
In the QDHA screening and quantum effects are both taken into account in
a first approximation. The inclusion of further effects as Hartree-Fock
etc.\ by means of Padé approximations makes no principal difficulties
but increases the numerical efforts. We will show that in the region 
of partial ionization and dissociation the QDHA yields a reasonable description of 
ionization/dissociation phenomena. The interactions of the neutral
components are taken into account by hard core approximation including
effects of reduced volume \cite{ebetal91,beuleal99}.

We will discuss now these contributions in more detail:
\begin{itemize}

\item the ideal free energy of electrons \( f_{e}^{id} \). We use Fermi-Dirac
statistics including excluded volume corrections representing the free energy in the following way
\cite{ebetal91}:
\begin{equation}
f_{e}=n_{e}k_{B}T\cdot z\!\left( \frac{1}{2} \frac{n_e \Lambda^{3}_{e}}{1-\eta_0}\right).
\end{equation}
The function $z(x)$ which is due to Zimmermann \cite{ebetal91} interpolates between the low density limit:
\( z(x) \approx \ln x-1 \) and the high density limit:
\( z\left( x\right) \propto x^{\frac{2}{3}} \). Further $\Lambda_{e}=h/\sqrt{2\pi m_e k_B T}$ 
is the thermal wave length of the electrons, correspondingly we will
use next the thermal wave lengths of ions, atoms and molecules ($\Lambda_{i}$, $\Lambda_{a}$, $\Lambda_{m}$). 
The excluded volume factor $\eta_0 < 1$ expresses that some part of the total volume 
$V$ which is occupied by atoms and molecules is not accessible to the electrons. The accessible volume
$(1 - \eta_0) V$ is smaller than the total volume and the effective densities are higher.
A strict theory for $\eta_0$ is not yet available. We will use here 
\begin{equation}
\eta_0=\frac{4}{3}\pi R_0^3(n_a+2n_m)
\end{equation}
as the definition for the packing parameter with an estimate for the atomic radius $R_0=0.78\mbox{Å}$.
Effects of a magnetic field are included by the replacement \cite{ebstor00}
\begin{equation}
n_e\Lambda_e^3\to n_e\Lambda_e^3\frac{\tanh(x_e)}{x_e},
\label{eq:replace-fe}
\end{equation}
where $x_e=\hbar\omega^e_c/2k_BT$ and $\omega^e_c=e B/m_e$.

\item the ideal free energy of the ions \( f_{i}^{id} \) which is given
by the classical term
\begin{equation}
f_{i}^{id}=n_{i}k_{B}T\left[ \ln \left( n_{i}\Lambda_{i}^{3}\right) -1\right]- n_i k_B T \ln(1-\eta_0). 
\end{equation}
The second term describes the excluded volume effect for the classical bare protons \cite{ebetal91}. 
The parameter $x_i=\hbar\omega^i_c/2k_BT$ ($\omega^i_c=e B/m_i$) is very small $x_i\ll1$ for small $B$. Therefore the magnetic field effects can be neglected for protons.

\item the ideal atom contribution: \( f_{a}^{id} \). We assume Boltzmann statistics
i.e.
\begin{equation}
f_{a}^{id}=n_{a}k_{B}T\left[ \ln \left( \frac{n_{a}\Lambda_{a}^{3}}{\sigma_{a}}\right) -1\right]
\end{equation}
including the internal states by a Brillouin-Planck-Lar­kin partition
function \cite{ebetal91}
\begin{equation}
\sigma_{a}(\xi)=\sum_{n=1}^{\infty}n^{2}\left\{ e^{\left(\frac{\xi}{2}\right)^2\frac{1}{n^{2}}} - 1 - \left(\frac{\xi}{2}\right)^2\frac{1}{n^{2}}\right\},
\end{equation}
where $\xi=2\sqrt{\mathrm{Ry}/k_BT}$ with the ionization energy of hydrogen $\mathrm{Ry}\simeq13.598\mathrm{eV}$. Including the magnetic field we have to replace the atomic partition function $\sigma_a$ by an effective partition function,
\begin{equation}
\sigma_a(T)\to\sigma_{\mathrm{eff}}(T,B),
\label{eq:replace-sigma}
\end{equation}
which reads \cite{ebstor00}
\begin{equation}
\sigma_{\mathrm{eff}}(T,B)=\sigma_a(\xi_e)+\frac{x_e^2}{24}\left[\sigma_B(\xi_e)+\frac{\xi_e^4}{192}\left(1+\frac{\pi^2}{3}\right)\right]
\end{equation}
with the interaction parameter $\xi_e=e^2/4\pi\epsilon_0k_BT\lambda_e$ ($\lambda_e=\hbar/\sqrt{2 m_ek_BT}$ is the thermal wave length for the relative motion of electrons and protons (with an infinit proton mass)) and \cite{ebstor00}
\begin{eqnarray}
\sigma_B(\xi)&=&\sum_{n=1}^{\infty}2n^2(1+n^2)\left[e^{\left(\frac{\xi}{2}\right)^2\frac{1}{n^2}}-1-\left(\frac{\xi}{2}\right)^2\frac{1}{n^2}\right. \nonumber \\
&&\left.{}-\frac{1}{2!}\left(\frac{\xi}{2}\right)^4\frac{1}{n^4}\right]-\sum_{n=1}^{\infty}n^4(5+7n^2)\left(\frac{2}{\xi}\right)^2 \nonumber \\
&&{}×\left[e^{\left(\frac{\xi}{2}\right)^2\frac{1}{n^2}}-1-\left(\frac{\xi}{2}\right)^2\frac{1}{n^2}-\frac{1}{2!}\left(\frac{\xi}{2}\right)^4\frac{1}{n^4}\right. \nonumber \\
&&\left.{}+\frac{1}{3!}\left(\frac{\xi}{2}\right)^6\frac{1}{n^6}\right].
\end{eqnarray}

\item the molecular contribution: \( f_{m}^{id} \). We will use the representation
\cite{beuleal99}
\begin{equation}
f_{m}^{id}=n_{m}k_{B}T\left[ \ln \left( \frac{n_{m}\Lambda_{m}^{3}}{\sigma _{m}}\right) -1\right] +f_{m}^{vib}+f_{m}^{rot}
\end{equation}
including the internal molecular and atomic states,
\begin{equation}
\sigma_{m}=\sigma_{a}^{2}\exp\left( \frac{4.746\mathrm{eV}}{k_{B}T}\right),
\end{equation}
and the contributions of vibrational and rotational sta­tes \cite{fowler}
\begin{eqnarray}
f_{m}^{vib} & = & n_{m}k_{B}T\left[\ln \left( 1-e^{-\frac{T_{v}}{T}}\right)+\frac{T_{\nu}}{2T}\right],\\
 &  &  T_{v}=6210\mathrm{K} \nonumber \\
f_{m}^{rot} & = & n_{m}k_{B}T\left[ \ln \frac{T_{r}}{T}-\frac{T_{r}}{3T}-\frac{1}{90}\left( \frac{T_{r}}{T}\right) ^{2}\right] ,\\
 &  & T_{r}=85\mathrm{K}.\nonumber 
\end{eqnarray}

\item the Debye-Hückel contribution \( f_{DH} \). Using here the so-called
\( \frac{\Lambda}{8} \)-approximation \cite{ebetal91} we arrive at the formula
\begin{equation}
f_{DH}=-k_{B}T\frac{\kappa^{3}}{12\pi }\tau \left(\kappa\,a(T) \right).
\end{equation}
Here the inverse Debye-radius is defined by
\begin{equation}
\kappa =\sqrt{\frac{2n_{i}e^{2}}{\varepsilon_{0}k_{B}T}}
\end{equation}
and the effective thermal electron radius by
\begin{equation}
a(T) = \frac{\Lambda_e}{8}
\end{equation}
where $\Lambda_e$ is the electron thermal wave length defined above.
Further the \( \tau  \)-function has the standard Debye-Hückel form \cite{ebkrkr76}:
\begin{equation}
\tau(x) = \frac{3}{x^{3}}\left[ \ln \left( 1+x\right) -x+\frac{x^{2}}{2}\right].
\end{equation}
The advantage of the QDHA is that magnetic field effects may easily taken into account by a factor \cite{ebstor00}
\begin{equation}
a(T)\to a(T,x_e)=a(T)\cdot\left[1-\frac{1}{48}x_e^2\right].
\label{eq:replace-a}
\end{equation}
The magnetic field reduces the effective diameter. This reflects the fact that the magnetic field localizes the particles perpendicular to the field.

\item the hard-core contribution \( f_{hs} \). For this term we use the
Carnahan-Starling approximation 
\cite{ebetal91}:
\begin{equation}
f_{hs} = k_{B}T\frac{\left( n_{a}+2n_{m}\right) \left( 4\eta -3\eta ^{2}\right) }{\left( 1-\eta \right) ^{2}}.
\end{equation}
The volume fraction of the neutrals is here defined by the effective packing parameter
\begin{equation}
\eta =\frac{4\pi }{3}R^{3}\left( n_{a}+2n_{m}\right).
\end{equation}
Following the work of Juranek and Redmer \cite{jure00} we fix the effective packing radius of atoms at a mean value $R=0.37\mbox{Å}$. 
This is of course a rough approximation
which might be refined by taking into account different radii for the atoms and for the 
molecules and by introducing temperature-dependent radii based on fluid variational theory \cite{jure00}.

\end{itemize}

\section{Discussion of the approximations and influence of parameters}

For this section we neglect the magnetic field, i.e.\ we assume $B=0$. We are interested in partial ionization and dissociation, i.e.\ in the region of the density-temperature plane where electrons, ions, atoms, and molecules exist. In this region the electrons may show quantum effects but the ions, atoms, and molecules are still classical.

\begin{figure}

{\centering \resizebox*{1\columnwidth}{!}{\includegraphics{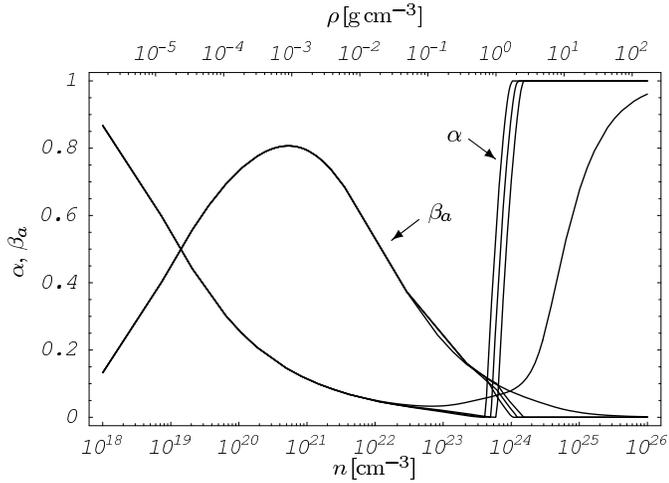}} \par}
\caption{\label{fig:AlphaBeta-ExclVol} Degree of Ionization and dissociation for different radii $R_0$ at a temperature of $T=2\cdot 10^4\mathrm{K}$. From right to left: $R_0=0$, $R_0=0.73\mbox{Å}$, $R_0=0.78\mbox{Å}$, and $R_0=0.83\mbox{Å}$}
\end{figure}

The excluded volume effect, i.e.\ the effect that the electrons and ions cannot penetrate into the neutral particles, are taken into account in the ideal contributions. The excluded volume factor $\eta_0$ is defined above with a fixed radius $R_0$. If this effect is not considered we find a lower ionization rate $\alpha$ at higher densities, shown in figure \ref{fig:AlphaBeta-ExclVol}. 

A small difference of the radius $R_0$ has an influence on the degree of ionization and dissociation, see figure \ref{fig:AlphaBeta-ExclVol}. For larger radii, $\alpha$ begins to increase at smaller densities since the effective volume $V_{\mathrm{eff}}=(1-\eta_0)V$, which is available for the electrons, decreases with increasing $\eta_0$ and thus increases the effective electron density $n_e^{\mathrm{eff}}=n_e/V_{\mathrm{eff}}$. Therefore the system tends to higher degrees of ionization. The transition between low and complete ionization takes place in a very narrow density range. For high densities the excluded volume effect has to be considered and has still to be refined maybe by taking into account temperature dependencies of the radius \cite{jure00} and by introducing occupation probabilities \cite{sach91,po96,chpo98}.
\begin{figure}

{\centering \resizebox*{1\columnwidth}{!}{\includegraphics{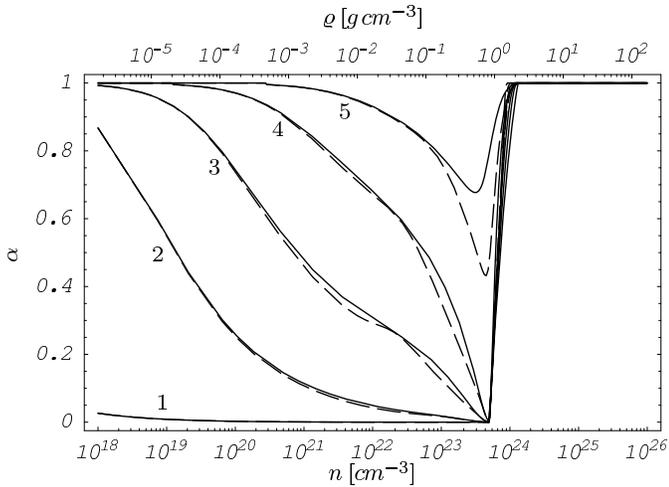}} \par}
\caption{\label{fig:AlphaDiffT} Degree of Ionization as a function of the total proton density for different temperatures: (1) $T=10^4\mathrm{K}$, (2) $T=2\cdot 10^4\mathrm{K}$, (3) $T=3\cdot 10^4\mathrm{K}$, (4) $T=5\cdot 10^4\mathrm{K}$, (5) $T=10^5\mathrm{K}$. The solid line represents the Calculation with the QDHA and the dashed line with Padé approximation.}
\end{figure}

The nonideal effects of the free charged particles can be described more correctly by means of Padé approximation (see e.\ g.\ \cite{ebri85,steb98,stbl00}). Figure \ref{fig:AlphaDiffT} shows the ionization and dissociation rate as a function of the proton density calculated by using the QDHA and by the Padé formulae given in \cite{steb98}. For the density range of $10^{22}-10^{25}\mathrm{cm}^3$ and the temperature range $1\cdot10^4-5\cdot10^4\mathrm{K}$ we find only small deviations of the QDHA from the Padé approach. For higher temperatures above $10^5\mathrm{K}$ the deviations are much larger and the QDHA is less accurate. In the following we will restrict our study to temperatures much below $10^5\mathrm{K}$, therefore it is justified to use further the QDHA to simplify the formulae and to reduce calculation efforts.

\section{The geometry of the free energy landscape}

\begin{figure}
{\centering \resizebox*{0.6\columnwidth}{!}{\includegraphics{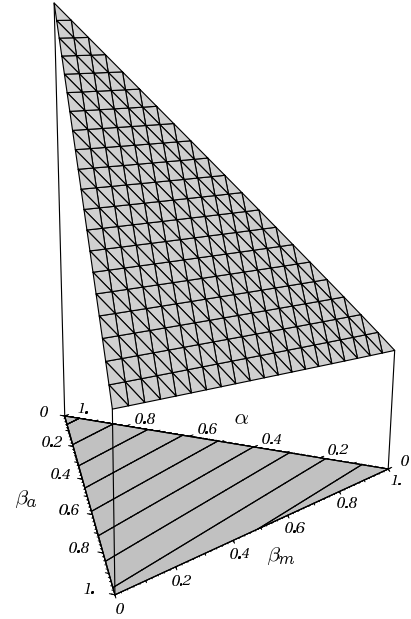}} \par}
\caption{\label{fig:FreeE-300K-2E19}Free energy density
at \protect\( T=300\mathrm{K}\protect \), \protect\( n=2\cdot 10^{19}\mathrm{cm}^{-3}\protect \) represented on the simplex $\alpha$, $\beta_a$, $\beta_m$.
The minimum of the free energy is at the corner $\alpha=0$, $\beta_a=0$, $\beta_m=1$,
i.e. in equilibrium we find practically only molecules.}
\end{figure}

\begin{figure}
{\centering \resizebox*{0.8\columnwidth}{!}{\includegraphics{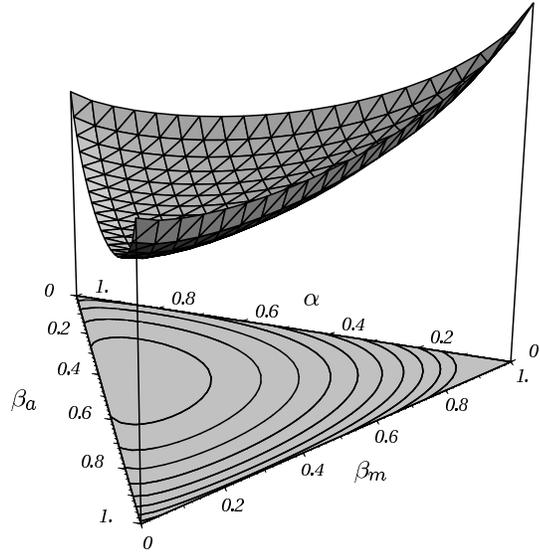}} \par}
\caption{\label{fig:FreeE-40000K-1E22}Free energy density 
at $T=4\cdot10^{4}\mathrm{K}$, $n=10^{22}\mathrm{cm}^{-3}$.
The minimum is found in a region where electrons, ions, atoms and
molecules are present.}
\end{figure}

On the basis of the formulae given in section \ref{sec:ContribFreeE} we represented
the geometry with respect to \( f\left( \alpha ,\beta \right)  \)
at fixed \( T \), \( n \) and starting with $B=0$ (figures \ref{fig:FreeE-300K-2E19}, \ref{fig:FreeE-40000K-1E22}).
We used the variables $\alpha$, $\beta_a$, and $\beta_m$ which allow a representation on a simplex according to eq.\ (\ref{eq:sumabc}).
The minimum of the free energy corresponds to the stationary
degrees of ionization and dissociation. The form of the landscape around the minimum 
determines the thermal fluctuations.

Depending on the values of temperature and proton density we find
different locations of the minimum. At \( T=300\mathrm{K} \), \( n=2\cdot 10^{19}\mathrm{cm}^{-3} \)
we observe a minimum in the corner $\alpha=0$, $\beta_a=0$, $\beta_m=1$, meaning that only
molecules exist (figure \ref{fig:FreeE-300K-2E19}). At different temperature
and proton density, e.g. $T=4\cdot10^4 \mathrm{K}$, $n=10^{22}\mathrm{cm}^{-3}$
we observe a minimum in the center of the simplex. This
corresponds to a situation where free electrons and ions and bound
states (atoms and molecules) exist under the same conditions (figure \ref{fig:FreeE-40000K-1E22}).

\begin{figure}
{\centering \resizebox*{0.9\columnwidth}{!}{\includegraphics{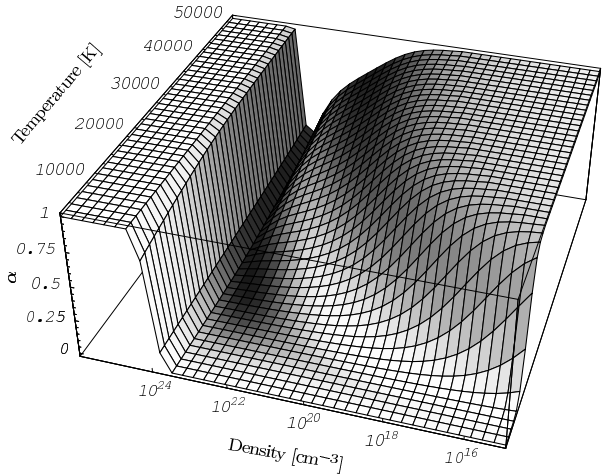}} \par}

\caption{\label{fig:AlphaVariance3D}The degree of ionization $\alpha$ together with the dispersion of $\alpha$ as a function of the proton density and the temperature. The coding is such that light regions correspond to small dispersion
and dark regions correspond to large dispersion.}
\end{figure}

With the procedure described above we calculated the stationary degrees
of ionization and dissociation for a lattice of points in the temperature-density-plane.
In this way we obtained the degree of ionization as a function of
density and temperature (figure \ref{fig:AlphaVariance3D}).

%
%
%

\begin{figure}
{\centering \resizebox*{0.9\columnwidth}{!}{\includegraphics{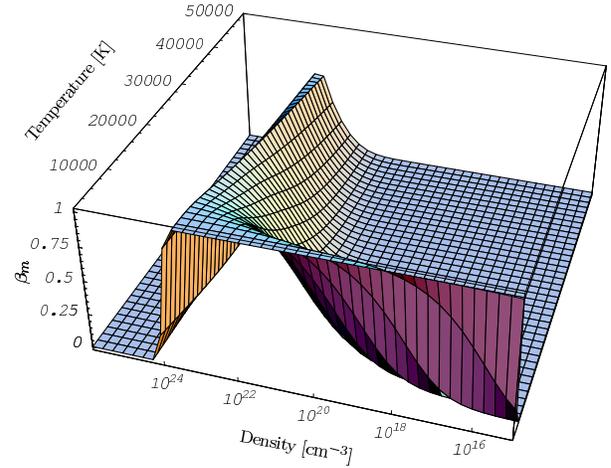}} \par}

\caption{\label{fig:Beta3D}The fraction of protons bound in molecules \protect\( \beta _{m}=\left( 1-\alpha \right) \left( 1-\beta \right) \protect \)}
\end{figure}

We observe full ionization (\( \alpha =1 \) on a plateau in the
upper right corner and at the left edge of the diagram) at very high
densities and in the region of lower densities but high temperatures.
At sufficiently low density we find only bound states when the temperature
is low (\( \alpha =0 \) in a region beginning at the lower edge of
the diagram). In the intermediate region of proton densities we find
a valley of bound states stretching over the whole temperature range
with small ionization taking place only at very high temperatures
\cite{ebkrkr76}.

Now we will look at the formation of molecules (figure \ref{fig:Beta3D}). We observe similar features as above in the corresponding parameter
regions. At very high densities (left edge) no molecules exist as
well as in the region of lower densities and high temperatures (upper
right corner). When the density is sufficiently low and the temperature
is low as well we find that the bound states observed in figure \ref{fig:AlphaVariance3D}
are practically only made up of molecules (\( \beta _{m}\approx 1 \)
at the lower edge). In the intermediate density region we find a mountain
of molecules (\( \beta _{m}=1 \)) and a decline of the degree of
dissociation with rising temperature.

\section{Fluctuations according to Boltzmann-Gibbs}

Let us first consider the general schema of the theory of isothermal
fluctuations at given temperature \( T \) \cite{ll,klim95}. We follow
Klimontovich who developed the general fluctuation theory of Boltzmann
and Gibbs \cite{klim95}. In the general case we study a set of intrinsic parameters
$A(X,a)=(A_1(X,a),\ldots,A_n(X,a))$ depending on dynamical variables \( X \) and
external parameters $a$ and define a conditional free energy
$F(a,T|A)$ which is the free energy at fixed values $A=A(X,a)$.
The intrinsic parameters $A(x,a)$ do not need to have any thermodynamic functions 
which correspond to them. The equilibrium value of the free energy is $F(a,T)$. Then according
to the Boltzmann-Gibbs principle the probabilities of fluctuations of $A$ 
are given by
\begin{equation}
\label{eq:probdist}
p(A|a,T)=\exp \left[ \frac{F(a,T)-F(a,T|A)}{k_BT}\right].
\end{equation}
In our case the dynamical variables are \( A=\left\{ \alpha ,\beta \right\}  \),
for \( F \) we choose the free energy per proton \( \phi =f/n \)
with the equilibrium value \( \phi _{0}=f_{0}/n \) at fixed temperature
\( T \).

Then according to the Boltzmann-Gibbs principle the probability distribution
of isothermal fluctuations reads 
\begin{equation}
\label{eq:prob-dist}
p\left( \alpha ,\beta | n,T\right) =\frac{e^{\frac{\phi _{0}-\phi \left( \alpha ,\beta \right) }{k_{B}T}}}{\sum _{\alpha ,\beta }e^{\frac{\phi _{0}-\phi \left( \alpha ,\beta \right) }{k_{B}T}}}.
\end{equation}
An illustration of the probability distribution
\begin{figure}
{\centering \resizebox*{0.9\columnwidth}{!}{\includegraphics{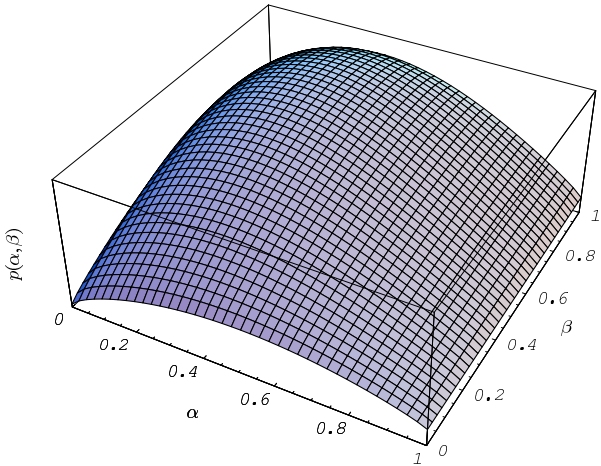}} \par}

\caption{\label{fig:E-Phi-30000K-1E23}The probability distribution in the
\protect\( \alpha \protect \), \protect\( \beta \protect \)-plane
at $T=3\cdot10^4\mathrm{K}$ and $n=10^{22}\mathrm{cm}^3$ }
\end{figure}
 calculated for a lattice of 50x50 points on the \( \alpha ,\beta - \)plane
is shown in figure \ref{fig:E-Phi-30000K-1E23}.

Further we show in figure \ref{fig:AlphaVariance3D} the degree of
ionization together with the dispersion of the degree of ionization
coded in gray levels.
%
%
 The diagram was obtained with the following procedure: For a given
temperature and proton density the stationary degrees of ionization and dissociation
(\( \overline{\alpha },\overline{\beta } \)) were calculated. Then
we additionally fixed the degree of dissociation at its stationary
value \( \beta =\overline{\beta } \) and calculated the probability
$p(\alpha)$ to find arbitrary values for the degree
of ionization. The dispersion of the degree of ionization $\overline{(\delta\alpha)^2}=\overline{(\alpha-\bar{\alpha})^2}$ was then
calculated as the root of the  mean quadratic deviation from the stationary value
$\overline{\alpha}$.

We observe low dispersion (light) at very high densities due to the
hard-sphere contribution. High dispersion (dark) is observed in two
transition regions, one when ionization processes start taking place
at intermediate densities and temperatures (just above \( \alpha =0 \)),
the other at the transition to complete ionization (just below \( \alpha =1 \))
where the remaining bound states break up.

Inside the simplex we assume that the probability distribution function $p(\alpha,\beta=\bar{\beta})$ is an one dimensional gaussian distribution 
for values $\alpha$ inside the borders and for small 
fluctuations $\delta\alpha = \alpha - \bar{\alpha}$
\begin{equation}
p(\alpha) = C \exp \left[-\frac{1}{2k_BT}\left(\frac{\partial^2\phi}{\partial\alpha^2}\right)_{\alpha=\bar{\alpha}}\left(\alpha-\bar{\alpha}\right)^2\right]
\end{equation}
where $\bar{\alpha}$ is the equilibrium value and therefore the most probable value. The constant $C$ is determined from the normalization. The dispersion in the fluctuations is given by
\begin{equation}
\overline{(\delta\alpha)^2}=k_BT \left[\left(\frac{\partial^2\phi}{\partial\alpha^2}\right)_{\alpha = \bar{\alpha}}\right]^{-1}.
\end{equation}
We see that the second deviation of $\phi(\alpha,\beta=\bar{\beta})$, which is the curvation of the free energy landscape 
determines the dispersion $\overline{(\delta\alpha)^2}$. Near to the corner of the simplex the distribution is highly non-gaussian and highly unsymmetrical. In this case the border of frequent fluctuations (the range of the dispersion) may be estimated by finding the roots of $p(\alpha)\simeq p(\alpha=\bar{\alpha})/\sqrt{e}$. For gaussian distributions this criterion agrees with the one given above. 

In density-temperature regions where the minimum of the free energy is developed very well e.g.\ if the minimum is located in a corner of the simplex (see e.g.\ figure \ref{fig:FreeE-300K-2E19}) we have small dispersions. Is the free energy landscape flat e.g.\ if the minimum is located inside the simplex (see e.g.\ figure \ref{fig:FreeE-40000K-1E22}) we have large dispersions.

\begin{figure}
{\centering \resizebox*{1\columnwidth}{!}{\includegraphics{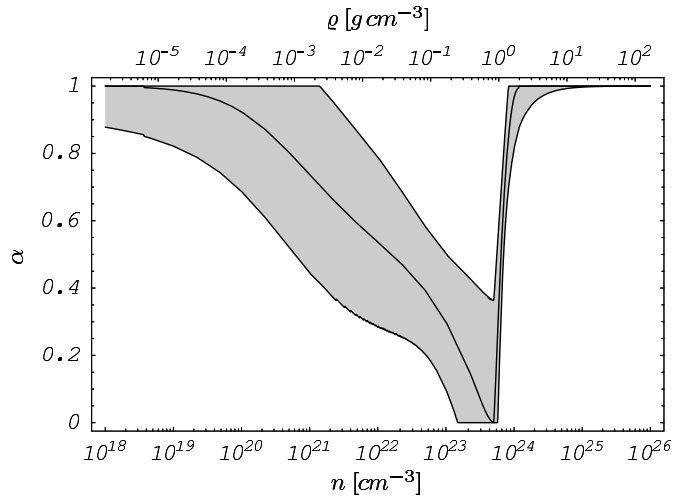}} \par}

\caption{\label{fig:AlphaVarianz2D-40000K}Degree of ionization as a function of the proton density $n$ with the dispersion of the fluctuations of $\alpha$ at $T=4\cdot10^4\mathrm{K}$.
}
\end{figure}

To further illustrate this, we show the dispersion of the degree of
ionization at fixed temperature, corresponding to a cut through the
diagram in figure \ref{fig:AlphaVariance3D} parallel to the \( n \)-axis.
The dispersion is indicated by error bars, which are cut at the physically
impossible values \( \alpha >1 \) and \( \alpha <0 \) (figure 
\ref{fig:AlphaVarianz2D-40000K}). Again the dispersion is small at
high and low densities and is high near the transition \( \alpha =1\Rightarrow \alpha <1 \)
where bound states emerge.

\section{Magnetic field effects}

\begin{figure}
{\centering \resizebox*{\columnwidth}{!}{\includegraphics{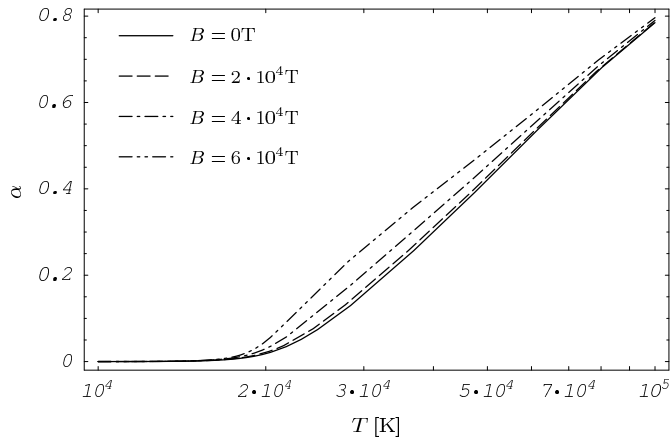}} \par}
\caption{\label{fig:AlphaOverT-DiffB} Degree of ionization and dissociation for different magnetic field strengths at a proton density $n=10^{23}\mathrm{cm}^{-3}$.}
\end{figure}

\begin{figure}
{\centering \resizebox*{1\columnwidth}{!}{\includegraphics{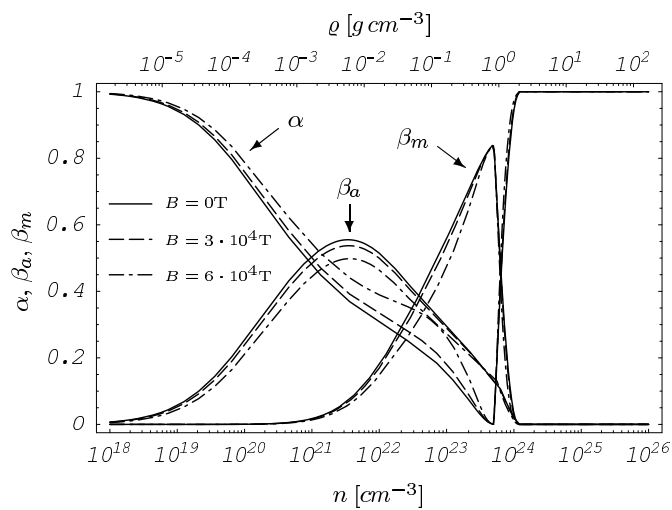}} \par}
\caption{\label{fig:AlphaBeta-DiffB} Degree of Ionization and dissociation for different magnetic fields at a temperature $T=3\cdot10^4\mathrm{K}$.}
\end{figure}

We want to discuss now the influence of a constant uniform magnetic field on the ionization and dissociation equilibrium. We have to carry out the replacements (\ref{eq:replace-fe}), (\ref{eq:replace-sigma}), and (\ref{eq:replace-a}). These formulae equations are valid for weak magnetic fields, i.e.\
\begin{equation}
\frac{x_e^2}{48} \ll 1.
\end{equation}
In figure \ref{fig:AlphaOverT-DiffB} the degree of ionization and dissociation for various magnetic field strengths are plotted over the temperature. We find at a fixed density that the degree of ionization of a plasma in a magnetic field is higher compared to the field free case and increases with the field strength. For temperatures higher than $10^5\mathrm{K}$ and weak magnetic fields there are no differences in the ionization degree. Further we see in figure \ref{fig:AlphaBeta-DiffB} that the curves deviate in the mid-density range where atoms and molecules exist up to the density where the transition to complete ionization takes place.

\section{Conclusions}

Based on a detailed study of the free energy landscape we developed
a new description of ionization and dissociation phenomena. The relatively
simple approximation QDHA used for the interaction term in the free energy 
allows the study of the dispersion of the degrees of ionization and 
dissociation and of magnetic field effects.
The QDHA may be replaced by more refined approximations as Padé approximation. 
We have shown that
the study of free energy landscapes in combination with the
fluctuation theory yields relevant
information on dispersion of the degrees of ionization and dissociation. 
The natural dispersion is usually in the range of $5-10\%$ except on the borders. 
This has consequences for many experimental quantities as e.g.\ conductivity 
and optical properties. Since the conductivity is in first approximation 
proportional to $\alpha$, i.e.\ $\sigma\sim\alpha\,\sigma_0$, we may conclude that the thermal 
dispersion of the plasma conductivity $\sigma$ has nearly the same shape as the 
dispersion of $\alpha$. We note that the thermal fluctuations of $\alpha$, $\beta$, $\sigma$ 
etc.\ are in general not of Gaussian character what is due to the existence 
of borders as e.g.\ $0\leq\alpha\leq1$. Studying the influence of magnetic fields we 
have shown that a remarkable influence on the degree of ionization starts 
at $B\sim10^4\mathrm{T}$, the effects increases with the magnetic field strength. 
At fields $B\gtrsim5\cdot10^4\mathrm{T}$ the degree of ionization and also the conductivity 
and other related quantities are substantially higher than for the field free case,
the present approach however is no more valid at these extremely high fields. \\
Recent experimental work on dense hydrogen concentrates mainly on Hugoniots \cite{knetal01}.
The approach presented here is restricted to temperatures $T > 10000 K$, therefore it 
cannot be used for the calculation of the low-temperature parts of Hugoniots. An extension
taking into account recent results on the low temperature EOS of dense hydrogen \cite{ebetal91}
is in progress.

\begin{acknowledgement}
The authors thank Stefan Hilbert, Hauke Juranek, Ronald Redmer, Gerd Röpke, Manfred Schlanges and Werner Stolzmann for many helpful discussions. The authors W. Ebeling and H. Hache profited very much from a stay at the Rostock University.
\end{acknowledgement}
\bibliographystyle{unsrt}
\bibliography{Ebhasp}

\end{document}